# Qualitative Research in an Era of AI: A Pragmatic Approach to Data Analysis, Workflow, and Computation

**Authors**

Corey M. Abramson, Department of Sociology and Center for Computational Insights into Inequality and Society, Rice University, Houston, Texas, United States

Tara Prendergast, Department of Sociology, University of Arizona, Tucson, Arizona, United States

Zhuofan Li, Department of Sociology, Virginia Tech, Blacksburg, Virginia, United States

Daniel Dohan, Philip R. Lee Institute for Health Policy Studies and Medical Cultures Lab, University of California San Francisco, San Francisco, California, United States

**Abstract**

Computational developments—particularly artificial intelligence—are reshaping social scientific research and raise new questions for in-depth methods such as ethnography and qualitative interviewing. Building on classic debates about computers in qualitative data analysis (QDA), we revisit possibilities and dangers in an era of automation, Large Language Model (LLM) chatbots, and 'big data.' We introduce a typology of contemporary approaches to using computers in qualitative research: *streamlining workflows*, *scaling up projects*, *hybrid analytical methods*, the *sociology of computation*, and *technological rejection*. Drawing from scaled team ethnographies and solo research integrating computational social science (CSS), we describe methodological choices across study lifecycles, from literature reviews through data collection, coding, text retrieval, and representation. We argue that new technologies hold potential to address longstanding methodological challenges when deployed with knowledge, purpose, and ethical commitment. Yet a pragmatic approach—moving beyond technological optimism and dismissal—is essential given rapidly changing tools that are both generative and dangerous. Computation now saturates research infrastructure, from algorithmic literature searches to scholarly metrics, making computational literacy a core methodological competence in and beyond sociology. We conclude that when used carefully and transparently, contemporary computational tools can meaningfully expand—rather than displace—the irreducible insights of qualitative research.

# Acknowledgements

We would like to thank Martín Sánchez-Jankowski, who offered feedback on the initial proposal and support to 'update' he and one of the authors' (Dan Dohan) classic 1998 ARS article on computing and ethnography. We extend gratitude to Claude Fischer and Ronald Breiger for their support and discussions on this work. We also thank Mary Waters, the ARS editorial team, and anonymous reviewers for useful feedback on this project. We acknowledge the generous feedback provided by Kieran Turner, Andy McCumber, Alissa Sideman, Kelsey Gonzalez, Alma Hernandez, Ignacia Arteaga, Melissa Ma, Brandi Ginn, Stephen Zaparista, Yuhan (Victoria) Nian, Jakira Silas, Zain Khemani and members of the Computational Ethnography Lab at Rice and Medical Cultures Lab at UC San Francisco. The authors are grateful to David Grusky and the full American Voices Project (AVP) team for their support in working with that large-scale qualitative data, the Qualitative Data Repository at Syracuse University for supporting work to verify associated papers, and the Russell Sage Foundation for facilitating opportunities to interact and share work. ASA events on "Trends in Mixed-Methods Research" and a panel on "Computational and Mathematical Approaches to Qualitative and Quantitative Data" organized by Laura Nelson provided additional inspiration.

This article draws on computational tools and workflows developed for the DISCERN study and prior ethnographic research. Example code, visualization templates, and sample data are publicly available at https://github.com/Computational-Ethnography-Lab/teaching (version: 2025-11-01), released under an open-source license for non-commercial use with attribution. Due to IRB restrictions, original transcripts and fieldnotes cannot be shared. Three large language models (Claude, GPT, Gemini) provided editing assistance for language clarity, citation coverage, and error detection. Models did not generate original content. All remaining errors are the authors' responsibility.

This research was supported in part by National Institute on Aging of the National Institutes of Health (NIA/NIH) award DP1AG069809 (Dohan PI). The content and views expressed are those of the authors and not of NIH.

The lead author previously used similar title language in a blog post: "A Silicon Cage? Qualitative Research in the Era of AI," Culture of Medicine, December 2023. The present article is substantially broader in scope, addressing computational methods across the qualitative research lifecycle rather than focusing primarily on AI-specific concerns in early qualitative data analysis software implementation.

Corresponding author: corey.abramson@rice.edu

# Introduction

Rapid computational developments, including the proliferation of 'artificial intelligence' (AI), increasingly shape contemporary life and sociological research (Bail 2024; Burrell & Fourcade 2021; Edelmann et al. 2020; Pugh 2024). The shifting technological landscape raises questions for traditional in-depth qualitative methods like participant observation and in-depth interviewing. Building on debates about the role of computers in qualitative data analysis following the 25th anniversary of early debates about this issue, we revisit classic concerns (Dohan & Sánchez-Jankowski 1998; Franzosi 1998; Mohr & Duquenne 1997) and discuss the possibilities as well as dangers of combining qualitative research with computation in an era of automation, AI chatbots, and 'big data.'

We do not attempt to predict whether computational developments will yield more good than harm—a task involving prophecy as much as review. Instead, we describe pragmatic uses of current tools. We outline decision points and tradeoffs, situating them within a broader sensibility that extends to computational technologies more broadly, and how they might be engaged sociologically.

First, we review and historicize methodological developments, noting that social scientists have used (and debated) computation in qualitative research for decades. New technologies rekindle longstanding concerns, such as the imposition of quantitative logics (Biernacki 2014; Burawoy 1998), distancing from data (Paula 2020), and the loss of contextual granularity (Grigoropoulou & Small 2022). New technologies also raise new challenges , such as AI hallucinations (chatbots generating false information with confidence) and algorithmic biases in how complex systems process people (Grossmann et al. 2023; O'Neil 2016).

Second, we turn to emergent scholarship at the intersection of qualitative research and computational social science (CSS). We examine a growing body of work that argues that contemporary computation can enhance, facilitate or expand qualitative research and analysis when used thoughtfully and with defined purpose. We examine five broad approaches to engaging with computational analysis and workflow in historically qualitative research (e.g. ethnography, in-depth interviewing, historical analyses)—*streamlining/augmenting workflow*, *scaling up* qualitative projects (e.g. Chandrasekar et al. 2024; Edin et al. 2024), *hybrid analytical approaches* involving mixed-methods (e.g. Abramson et al. 2018; Nelson 2020; Pardo-Guerra and Pahwa 2022), and the *sociology of computation* (e.g. Christin 2020). These approaches, alongside the *rejection* of technologies, differ in their aims and means of technology use, shaping strategies of engagement in varied ways.

Third, we illustrate these approaches through workflow examples and case studies from our research, demonstrating how computational tools can be integrated into qualitative studies more broadly. We describe how technologies can be used to process and analyze information like ethnographic fieldnotes, historical documents and interview transcripts. We also discuss how they shape project design even when not intended—e.g. how algorithms mediate our literature reviews—and identify further possibilities for analysis, publication, and data sharing. Detailing our work on a large-scale ethnographic study that combines human interpretation with computational procedures, we explain the workflow and how key aspects can be adapted for different aims, as well as single-researcher studies (Arteaga et al. forthcoming 2025; Li and Abramson 2025).

We conclude by noting that debates over the deployment and repurposing of analytical technologies are a perennial topic in the history of social science. Sociology has a deep tradition of pragmatic and sometimes subversive adaptations of available tools. This legacy runs from W.E.B. Du Bois's (1900) innovative use of data visualization to challenge racial inequality to C. Wright Mills's (1959) call to transform the filing-cabinet and data archive into a systematic engine for imaginative inquiry. Following Mills's insight that well-developed systems enable rigorous creativity, computational tools can be repurposed—with the sociological imagination—as tools for machine-readable datasets, flexible indexing, and archival posterity (Abramson 2024). A key contribution is to articulate a sociological approach that moves beyond the dualism of technological optimism or rejection to show how computation can be adapted to support longstanding qualitative aims. In so doing, we model navigating computational tools that can be both dangerous *and* generative, constraining *and* useful, in ways while maintaining core methodological commitments.

## Qualitative Analysis and Computing: Debates, Longstanding and New

Qualitative research encompasses methods that are not easily reducible to numerical data, such as ethnography and in-depth interviews (Small & Calarco 2022). These methods are essential for capturing meanings and occurrences that might otherwise be missed (Glaser & Strauss 1967; Lamont & Swidler 2014), illuminating inequalities (Burawoy 1998; Collins 2000), documenting sequences of behavior in situ (Jerolmack and Khan 2014; Sánchez-Jankowski 2002), and preserving ecological validity (Cicourel 1982; Gans 1999). Qualitative research draws on diverse traditions – from phenomenology to realism, intersectional feminism to grounded theory – that have distinct philosophical priors (Gong & Abramson 2020; Steinmetz 2005).

This diversity contributes to confusion around terms. The idea of 'qualitative coding'—tagging or indexing text data with a theme, concept, category, or variable – provides a useful example. While some critique coding as reductive categorization (Biernacki 2014) and others see it as central to theorizing (Strauss & Corbin 1990), we frame it as *indexing*: a systematic way to tag and retrieve text without supplanting or precluding deeper interpretation (Li and Abramson 2025). This aligns with a pragmatic, iterative approach common in many strains of contemporary sociology that integrate prior theory with emergent findings rather than adhering to a strict inductive or deductive logic (Abramson & Sánchez-Jankowski 2020; Deterding & Waters 2021; Lichtenstein & Rucks-Ahidiana 2023). Our approach uses indexing to identify patterns and navigate complexity without reducing text to codes alone.[1]

Despite the variation in approach, some challenges and imperatives are shared. In addition to the challenge of navigating definitional issues, qualitative scholars concerned with systematic inquiry face the same practical task of qualitative data analysis: how to examine and share information that cannot be reduced to numbers alone without compromising the purpose of collecting in-depth data in the first place (Abramson & Dohan 2015). Computational tools can and have been used in a variety of ways to accomplish this.

*Computational Tools for Qualitative Data Analysis*

[1] 'Coding' attracts varied critiques across research paradigms. However, computational tools offer value beyond traditional coding through visualization, systematic retrieval, and file management—approaches that accommodate diverse interpretive traditions.

The use of computational methods, or computation, for managing and analyzing qualitative data has a long history that predates generative AI (i.e. computer systems that generate new content by learning patterns from existing data).

For decades, researchers have used commercial software, home-made retrieval systems in markdown, notebooks and word processors, spreadsheets, and occasionally (computational) code to analyze and write about people. Early forms of Computer-Assisted Qualitative Data Analysis Software (CAQDAS) were designed to streamline labor-intensive tasks involved in managing text data such as ethnographic fieldnotes, interview transcripts, and historical documents (Silver 2018). These tools supported core processes such as aggregating, structuring, tagging/coding, memoing, and retrieving text from fieldnotes, transcripts, and documents (Dohan and Sánchez-Jankowski 1998). Early work by scholars including C. Wright Mills and W.E.B. Du Bois repurposed bureaucratic technologies such as filing cabinets, carbon-paper and folders to enable the sociological craft before the dawn of modern computing (Abramson 2024). While sociological applications have sometimes integrated text and computational modeling (Mohr and Duquenne 1997; Roberts, et al. 2022), most qualitative uses of computers—typically relying on Graphical User Interfaces (GUIs), the point-and-click software familiar to users of commercial programs like ATLAS.ti or NVivo—have emphasized reducing time burdens and reliance on memory, making established practices more efficient rather than fundamentally rethinking the possibilities of analysis and preservation.

## Definitions and Acronyms

- ***Computation*** here refers to the use of digital technologies in research. ***Computational tools*** are software, programming languages, and plugins used for computation (e.g., Python, R, specialized packages and toolkits, large language models, markdown editors, writing assistance plugins such as Grammarly).
- ***Computer-Assisted Qualitative Data Analysis Software (CAQDAS)*** is specialized software for qualitative data analysis (e.g., MAXQDA, NVivo, ATLAS.ti, Dedoose) that typically includes automation and AI features. CAQDAS is a type of computational tool.
- ***Artificial intelligence*** (AI) refers to technologies designed to mimic human performance on tasks that historically required human intelligence. This can include recognizing patterns, extracting text from .pdf files, classifying images, summarizing interviews, or generating synthetic content such as manipulated images or text. Some subfields commonly used in qualitative research workflows include ***machine learning (ML)*** for analyzing behaviors and cases, ***natural language processing (NLP)*** for parsing language data, and computer vision for analyzing images.
- ***Large language models (LLMs)*** – deep learning systems trained on mass-scale text data to predict and/or generate language (GPT is a commercial example) – are a subset of AI. While sometimes integrated into qualitative analysis, LLMs pose various methodological and ethical challenges as well as possibilities.

Over time computational tools used with qualitative data have expanded to include possibilities for quantification, visualization, and automation as well as emergent forms of new scholarship (Bjerre-Nielsen & Glavind 2022; Li & Abramson 2025; Peponakis et al. 2023). The current moment abounds with technologies that alter, extend or reconfigure capabilities for qualitative data analysis. Recent work demonstrates how purposeful machine learning can scale researcher-created codes (concepts, themes, categories) using inductive, deductive, and iterative approaches (Jarrahi 2025; Li et al. 2021; Nelson et al. 2021; Pardo-Guerra & Pahwa 2022; Peponakis et al.

2023). AI tools are being used to aid indexing of satellite imagery and historical texts (Law & Roberto 2025), transcribe and summarize audio, visualize and triangulate patterns in field-research (Abramson et al. 2018; Hanson and Theis 2024), identify cultural concepts embedded in large text datasets (Boutyline & Arseniev-Koehler 2025; Kozlowski & Evans 2025), support counterfactual checks and retrieval (Davidson 2024; Li et al. 2021; Rodriguez and Spirling 2022; Stuhler, Ton, and Ollion 2025). They are even being used to simulate human beings—creating computational models that generate synthetic data (e.g. new text based on prior data) or responses (Kozlowski and Evans 2025) based on ideal-typical personas – in order to produce computer-generated interview-like responses to explore what people might say, or consider, under specific circumstances.[2] Each has support and detracting positions, to varying degrees as social science fields grapple with how to integrate new technologies, a question at present not yet settled (Abramson 2024; Davidson 2024; Davidson & Karell 2025).

The technological developments of the AI era exacerbate existing concerns and present new challenges. In particular, the degree of automation and growth of complex systems under the umbrella term of AI, introduce new concerns and longstanding concerns about the use of computers in qualitative research.

### *Concerns and Critiques*

Even when computational tools are used to support conventional aims such as aggregation, organization, coding, and retrieval of in-depth text account, they (including CAQDAS) have faced persistent critiques, including:

- **The imposition of a singular logic.** Critics argue that computational tools impose a standardized logic on researchers that results in quantifying and structuring fluid, context-dependent data in ways that prioritize measurability rather than deep reading and interpretation (Biernacki 2014; Coffey et al. 1996). Biernacki (2014) argued that coding itself risks reducing text to variables, though others note contemporary 'codes' can function more like interactive hashtags that index text without replacing it (Li and Abramson 2025; Deterding and Waters 2021). This concern persists even as software now allows full reading and linking of complete documents—the core point being that tools (or their absence) shape our engagement with data even if they are flexible enough to allow both grounded theory and positivist inquiry.
- **Distancing from data.** Qualitative work depends on close, immersive engagement with source materials, often drawn from in-depth direct human interactions (Lareau and Rao 2016; Fine and Abramson 2020). Interfaces that decontextualize quotes or excerpts have been criticized for creating analytic and emotional distance. Even when human review remains central, some worry that insights from slow, manual iteration may be lost (Paulus & Marone 2024).
- **Decontextualization.** Computational methods risk stripping qualitative data of its meaning when context is reduced or ignored. An utterance or fieldnote gains its significance from a network of social, linguistic and historical context (Cicourel 1982; Geertz 2000). Context can vanish when data is quantified and separated from the textual representation, risking superficial or erroneous interpretations akin to mining for "statistical rather than sociological significance" (Bourdieu 1984; Lichtenstein & Rucks-Ahidiana 2023).

[2] Social science has long used abstractions without direct human subjects (agent-based models, formal theory), even though doing so has been less common with qualitative research. Simulations produce synthetic approximations that abstract from original meanings, in ways distinct from direct human accounts recorded as text. This review focuses on more traditional qualitative data: immediate observations, accounts and documents from people, though we discuss the new wave of simulation research in the appendix.

Newer concerns with the use of computing in qualitative analysis focus on the rise of large language models (LLMs), which have been used alongside or as a replacement for traditional techniques. These include:

- **Technical unreliability and bias.** LLMs often produce inconsistent results, shifting with small prompt changes or model updates, typically without transparency (Bisbee et al. 2024; Chen et al. 2024) They are prone to "hallucinations"—plausible but incorrect outputs—which may be underestimated given intuitive interfaces (Farquhar et al. 2024). Training corpora (text data used to construct the models) introduce algorithmic bias (Ashwin et al. 2025), while failures of algorithmic fidelity obscure data nuance (Amirova et al. 2024). Results are difficult to replicate absent validation standards (Abdurahman et al. 2025). LLMs struggle with irony, sarcasm, and cultural nuance without specialized tuning (Weber & Prietl 2021).
- **Social harms and flawed interpretation.** Without curation and alignment models can reproduce stereotypes and biases (Omiye et al. 2023; Tan & Celis 2019), misrepresent groups (Santurkar et al. 2023), and amplify particular viewpoints as 'common sense' (Bender et al. 2021). In health care and organizational psychology, algorithmic sorting by race worsens disparities (Hurd et al. 2024; O'Neil 2016). LLMs rely on exploitative labor for data training and curation, reinforce racial hierarchies (Bender and Hanna 2025; Benjamin 2019), and erode human roles (Pugh 2024; Zuboff 2019).When used as proxies for human interpretation, they risk reproducing normative narratives, fabricating data, and extending classification systems rooted in historical, racialized disparities (Büscher 2025; Crawford 2021; Farber et al. 2025; Hayes 2025; The Ghost in the Algorithm: Racial Colonial Capitalism and the Digital Age 2021b) ( Büscher 2025; Crawford 2021; Farber, Abramson, and Reich 2025; Hammer and Park 2021).
- **Security risks and corporate dependency.** Using third-party commercial LLMs to analyze sensitive data raises security and confidentiality risks when cloud-based chatbots handle qualitative material (Bail 2024; Chopra and Haaland 2023; Farber et al. 2025). The concentration of AI development in a few corporations fosters dependency and global inequality, while opaque training decisions and precarious labor arrangements sustain these systems (Burrell and Fourcade 2021; Spirling 2023). Deploying LLMs as "synthetic participants"—AI-generated personas that mimic human responses—potentially add concerns. These tools risk substituting algorithmic proxies for authentic voices, complicating consent, and potentially detaching data from its original context. Such practices raise questions about appropriation, accountability, and downstream data uses—particularly when exceeding participants' expectations or benefiting commercial platforms. Environmental costs of large-scale computation remain substantial, though commercial deployments, not scientific research, are the primary contributors (Bender et al. 2021; Kozlowski and Evans 2025; Strubell, Ganesh, and McCallum 2020).

The appendix summarizes some risks and mitigation strategies in a table.

*Moral Panic or Legitimate Concerns?*

Some concerns long associated with traditional CAQDAS, and applied to other computational tools, never fully materialized. Researchers have adapted such tools to a wide range of paradigms—from conventional science to post-modern critique—and while early platforms sometimes decontextualized excerpts, most contemporary software now enables reading full passages in context. Automation can assist with preliminary indexing, but in sociological analysis it does not substitute for interpretive reasoning or the application of complex concepts, making reduction a user choice rather than a built-in feature (Deterding and Waters 2021; Dohan and Sanchez-jankowski 1998; Li et al. 2021; Lichtenstein and Rucks-Ahidiana 2023). Meanwhile, both project scale and computational experimentation are increasing. Proponents of "scaling up" qualitative research argue this can support comparative or representative designs without losing contextual depth—a tension that parallels debates in big data research about the need to "scale down" to meaning (Abramson et al. 2018; Breiger 2015; Edin et al. 2024; Nelson 2020). Even solo researchers can now use machine learning to extend iteration or scale interpretive categories. As with word dictionaries or statistical tools, software is not inherently reductionist, but can be deployed to diverse analytic ends.

Responses to AI are part of broader social transformations and debates about emergent technologies . Some argue that fear of emerging technologies verges on moral panic, potentially derailing productive inquiry (Simon, Altay, and Mercier 2023; Spirling 2023; Yadlin and Marciano 2024). Others point to substantive concerns—especially with large-scale, commercial LLMs. The question is not whether risks exist, but whether they are tractable, with some researchers citing parallels in scientific computing and inductive statistics as precedent for responsible adaptation (Bail 2024; Davidson 2024). Pragmatic applications already exist: researchers can run open-source models on personal computers, customize them for specific research contexts, and analyze data without transmitting information to external servers (Abramson 2024; Li et al. 2021), and there is work on critical engagements with algorithmic infrastructure (Benjamin 2019).

While AI has expanded into everyday domains—from college essays to website blurbs—this broadening may obscure the distinctions among very different technologies used in and outside of social science (Bender and Hanah 2025). The concern is not just technical: science has advanced human understanding and well-being but has also been used to marginalize and harm. This applies not only to fields like biology and physics, but also to ethnography, statistics, and computing (Foucault 1977; Jasanoff 2004; Noble 2018). Compared to the 1998 article that inspired this work, computers are now inescapable in research: used for literature reviews, writing, data management, article proofing, and dissemination, making full avoidance nearly impossible. As Weber noted in his account of rationalization, once systems are institutionalized, opting out becomes not only difficult but itself a choice with consequences (Weber 1978).

Computing, like earlier scientific advances, may exacerbate inequality or help document and reduce it. Ethnography has served colonial expansion and the pursuit of justice; physics enabled both weaponry and renewable energy. Specific uses matter: the validity of critiques and the appropriateness of applications depend on particular implementations and contexts rather than computation in general. Employing local, open-source language models to check patterns in human-coded data differs markedly from using commercial chatbots as stand-ins for analysis. The former runs on secured systems, ideally fine-tuned for interpretive rigor and privacy; the latter sends potentially sensitive material to opaque third-party providers. Even among commercial tools, distinctions matter: browser-based chatbots differ from API integrations in terms of control, formatting, and data flow.[3] Both approaches fall under qualitative analysis and both involve AI—yet they differ sharply in flexibility, interpretive transparency, and data security. As with any analytic decision, these practices must be accounted for and justified. Computing may extend surveillance, but it may also enable researchers to analyze large volumes of qualitative data that were once accessible only to elite labs (Bail 2024). Our task is to determine how such tools can be used in ways that serve our values and scholarly aims.

[3] API refers to application programming interface, allowing communication between programs so data can be sent and received automatically between services. For instance, connecting visualization software to annotation tools. In our work, we use small, fine-tuned models on local university servers or secure desktop systems—without sending data externally. These differ from both GPT-style browser chatbots and API-based AI services, which require third-party trust. These distinctions—local, institutional-secure, versus commercial cloud platforms—carry implications for data governance and transparency.

Having established a few core premises for this article, we now turn to the range of ways scholars are engaging with emerging technologies to support, extend, or transform qualitative research and what exactly that means in practice.

# A Typology of Qualitative–Computational Engagement

Combinations of qualitative methods and computation are tremendously varied, reflecting the diversity of qualitative traditions and changing computational tools themselves. Following Merton's (1938) discussion of social structure and anomie—which theorizes how people understand goals (valued ends) and means (institutionalized possibilities) and adjust through strategies of conformity, innovation, ritualism, etc.—we organize qualitative-computational approaches into genres of engagement. These ideal-typical styles vary in whether they extend widely valued qualitative goals (understanding and interpreting human action) or expand toward new ones (predicting outcomes). They also vary in whether they rely on conventional means (e.g., software for extending codes) or require expanded computational techniques (e.g., using AI to create interactive simulations). This typology helps situate contemporary practices and clarify how scholars engage computation.

## Genres of Engagement

**Streamlining (extending goals, extending means)**
Many approaches use computational tools to extend existing qualitative practices without radically altering them. The goal is continuity and enhancement rather than expansion, with technology serving to improve efficiency or consistency in tasks such as coding, retrieval, and comparison. Sociological examples include flexible and context coding that involve basic automations in QDA software for medium to large datasets (Deterding & Waters 2021; Lichtenstein & Rucks-Ahidiana 2023), and human-hybrid machine-learning approaches that improve label consistency while maintaining human judgment in text classification (Li et al. 2021). These methods extend established qualitative workflows to facilitate or enhance interpretive analysis rather than replace it with numerical information. Recent analyses suggest that many works use new LLM technologies with conventional approaches—such as inductive coding, thematic analysis and the deductive application of concepts—though sometimes automating in ways that extend beyond streamlining (Abramson 2024; Chandrasekar et al. 2024; Than et al. 2025)

**Scaling Up (extending goals, expanding means)**
Other projects extend the goals of qualitative inquiry while expanding how they are achieved by focusing on how computational infrastructure can scale up the size, reach, and accessibility of qualitative data. These approaches align with open science initiatives that encourage transparency, sharing, and replication through expanded tools for de-identification, sharing, validation and reuse (Abramson et al. 2018; Freese et al. 2022; Kirilova & Karcher 2017).

Examples include the American Voices Project (2021), which collected over 1,500 interviews using stratified sampling with weights to approximate national representativeness (Edin et al. 2024). Comparative ethnographic projects have similarly used shared protocols and digital infrastructure to generate policy-relevant qualitative data (Bernstein and Dohan 2020), while others have restructured sampling and analysis for generalizability at scale (DeLuca, Clampet-Lundquist, and Edin 2016; Sykes, Verma, and Hancock 2018). Scaling does not necessarily alter the core goals of qualitative research—such as interpreting lived experience or observing behaviors in situ—but it does require computational and procedural expansion to manage the volume of 'big' qualitative data, moving beyond earlier large-scale efforts like Mass Observation or the Human Relations Area Files to contribute public insights into human life (Abramson and Dohan 2015; Karcher et al. 2021; Murphy, Jerolmack, and Smith 2021). [4]

Computational infrastructure becomes necessary—not merely preferable—for large-scale qualitative projects. Traditional QDA software often cannot handle the volume of data generated by team ethnographies or large interview studies, requiring researchers to turn to open-source packages in R or Python. Commercial QDA platforms present additional concerns: high costs that limit accessibility, opaque algorithms for text analysis that obscure analytical processes, and increasingly, AI features that may transmit data to external language models despite user agreements meant to address these risks.

**Hybrid approaches (expanded goals, expanded means)**
Hybrid approaches—rooted in sociology's pragmatic tradition and history of multi-method triangulation (Du Bois 1996; Small 2011)—reflect an expansion of goals to align with multi-method inquiry. This growing body of work combines qualitative interpretation with computational tools to expand qualitative research in more fundamental ways—aiming to reconfigure what is possible, or the types of questions that can be asked. This includes computational grounded theory for identifying themes (Nelson 2020), computational ethnography for charting patterns of speech and behavior within and across cases (Abramson et al. 2018), and the extended computational case method which examines data construction itself (Pardo-Guerra and Pahwa 2022). Each uses multiple methods to connect macro patterns and micro data. Such approaches, developed by scholars studying culture and inequality (Abramson et al. 2018; Nelson 2020), knowledge construction (Pardo-Guerra and Pahwa 2022), and health (Bernstein and Dohan 2020), allow researchers to examine larger datasets and connect multilevel patterns. They aim to bridge dualities between history and biography, objectivity and subjectivity, individual and group by showing both patterns and specific manifestations in micro-data (Breiger 1974; Du Bois 1996; Mills 1959; Mohr and Duquenne 1997).

Hybrid approaches may be combined with attempts at scaling using extant or prospective data (Abramson et al. 2024) or used on data of traditional scope to analyze it from multiple vantages in exploratory, confirmatory or explanatory ways (Garrett et al. 2019; Mische & Mart 2025).

---

[4] Scaling up also pushes us to revisit assumptions like "saturation." In large studies, the aim shifts from "no new themes" to heterogeneity, disconfirming cases, and broad conceptual or geographic coverage. Saturation—often defined as the point when no new codes or themes appear—has been widely critiqued, especially when used to justify small samples. Recent work suggests that conceptual saturation—adequate coverage of meanings—often requires ~50–100 interviews per group, far beyond the familiar 15–20 (Low 2019; Rahimi 2024; Saunders et al. 2017; Squire et al. 2024)(Low 2019; Saunders et al. 2017; Squire et al. 2024; Rahimi 2024). At scale, researchers subset by metadata (e.g., age, topic) and use automated tools to surface relevant segments (e.g., older adults, gig workers, people reporting pain). Coding therefore may prioritize coverage, contrast, and transparency over claims of completion; index codes organize material and do not require saturation.

**Studying computation itself**

Qualitative methods' uneasy relationship with computing is compounded by work that makes computing and digital life objects of inquiry in their own right (Burrell and Fourcade 2021). While social science has long studied sociotechnical systems, large-scale computing introduces new methodological challenges. Algorithms are often opaque—hidden by organizational secrecy, user illiteracy, and, in AI, a lack of interpretability even for experts (Burrell 2016). Studying algorithms forces ethnographers to shift across levels of analysis, from individuals engaging code to platform economies, and to reassemble diffuse networks of human and non-human actors (Cottom 2020; Rosa 2022; Seaver 2017). Even when not the focus, algorithms' ubiquity demands engagement: researchers may need to 'enroll' them in producing qualitative data (Christin 2020). For example, since recommendation systems increasingly shape social media, understanding how they translate data into user experience is now prerequisite for studying online activism (Earl & Kimport 2013), urban neighborhoods (Lane 2018), or the creator economy (Mears 2023). Researchers must likewise account for the extent and automation of content moderation (Roberts 2019; Mears, Birced, and Nguyen 2025). On labor platforms, algorithmic control makes architecture itself central to many service jobs, online and offline (Griesbach et al. 2019; Ravenelle 2019). Algorithms also mediate the visibility and circulation of scholarship—through platforms like Google Scholar and citation metrics—shaping what becomes legible in the academic field and how scholars respond beyond methodological choices (Pardo-Guerra 2022). These dynamics have led digital sociologists to integrate computational and mixed methods into domains traditionally reliant on qualitative data, in order to theorize human experience in an algorithmic society. Crucially, sociological engagement with AI must examine not just technical capabilities but the social contexts shaping AI's development and use—including funding sources, organizational contexts, the use of technology itself, and how the uneven uptake of these tools contributes to social inequalities (Ashery et al. 2025; Hayes 2025; Joyce et al. 2021).

**Rejection (rejected goals, rejected means)**

Finally, not all qualitative scholars embrace computational methods, even at the level of software for organization. Some reject computation outright, either on epistemological grounds or in defense of a humanist orientation. These perspectives argue that attempts to formalize, automate, or quantify qualitative work distort its interpretive aims and risk reinforcing technocratic or 'positivist' logics (Biernacki 2014; Burawoy 1998). This stance often represents a rejection of the 'social science turn' in favor of traditions that prioritize meaning, resistance, and humanism (Clifford & Marcus 1986). Rejection thus resists aligning qualitative research with computational or data-intensive science, which is perceived as the encroachment of a singular, technocratic logic or even the dangers of empiricism.

The typology is represented in table 1 below.

**Table 1: Genres of Qualitative – Computational Engagement in Research***

| Research Goals | | |
|---|---|---|
| **Technical Means** | *Extension* | *Expansion* |
| *Extension* | **Streamlining**<br>Using software more efficiently, without radically altering qualitative practices | **Sociology of Computation**<br>Defined by topical area<br><br>May involve new tools |
| *Expansion* | **Scaling Up**<br>Large-scale fieldwork<br>Team based interviews<br>Shared repositories | **Hybrid Approaches**<br>Mixed methods<br>In-depth text analysis and computational analysis together |

**Not shown: The 'Rejection' genre (rejecting computational means and/or goals) is discussed in the text but not shown in the table for space.*

Taken together, this typology highlights both commonality and divergence in how sociologists and other researchers engage computation in qualitative research. Some seek efficiency, others scale, others hybrid innovation, and some resist or critique computational tools altogether. These positions reflect deeper debates about what computation enables or limits in qualitative inquiry, how researchers should relate to these tools, and what epistemological tensions arise when merging computational and qualitative traditions (Schroeder et al. 2025). In the next section, we turn to practical workflows and case studies that illustrate how these tensions play out in applied settings. Our example reflects only one way of integrating computation. It is guided by specific values and project goals, and theories of method not offered as a universal model.[5]

We now turn to practical workflows and case studies that illustrate applications of technology to qualitative research and data analysis.

## Workflow and Uses of Computation in Qualitative Research (2025 Redux)

In the current moment of proliferating AI, the potential uses of computation in qualitative research spans the entire project lifecycle, from study design to dissemination. Much like the earlier adoption of CAQDAS (Coffey et al. 1996; Dohan and Sánchez-Jankowski 1998), current tools—from shared datasets like the American Voices project, open-source visualization resources, and commercial LLMs—can reshape practices, addressing or introducing challenges of rigor, interpretation, and transparency over the course of a project.

[5] Without an explicit theory of method, researchers risk embedding unexamined assumptions—such as viewing individuals as atomized actors or importing cultural prejudices—into their analytical frameworks. Such methodological opacity can invite the uncritical adoption of common sense or ideology as scholarly reasoning. Similarly, method without underlying theory can sever the connection between empirical observation and analytical constructs, producing topics, cases, and comparisons divorced from the social realities they purport to explain (Abramson and Gong 2020; Cicourel 1964), a dangerous form of abstracted empiricism that adds fails (Mills 1959).

We first provide an overview of workflow possibilities, illustrated with examples of current uses. Table 2 presents these applications in roughly chronological order (from project start to finish), distinguishing whether computational tools are broadly assistive (basic technologies predating contemporary applications, but involving a computer), automated (narrowly defined tasks, often integrating aspects of machine learning), or agentic (broader execution by computing systems that heavily involve LLMs). Rows can be thought of as phased with decision points in research, and columns as how computers may be integrated.

**Table 2: Workflow Stages and Computational Tools**

| | **Assistive** | **Automated** | **Agentic** |
|---|---|---|---|
| **Research Design** | Citation management<br>Project records<br>Project management<br>Version control | Readability checks<br>Data-assisted sampling<br>Simulating sample-size | Literature review |
| **Data Collection** | Participant/site tracking<br>Hyperlink field artifacts<br>E-consent capture<br>Digital diary or EMA<br>Survey apps<br>Cloud back-up/sync | Multi-media aggregation<br>Sensor or geospatial logging<br>Timestamping/geolocation<br>Live transcription | Adaptive or event based SMS prompts |
| **Data Processing** | Interview transcription<br>Transcript editing<br>File-format normalization<br>Data versioning<br>Data tracking | Scanned docs/images to text<br>A/V speech-to-text pipelines<br>Context/metadata auto-capture<br>Entity/event annotation layers<br>De-identification workflows<br>Batch uploads<br>Text into data frames/databases<br>Metadata tagging<br>Quality checks<br>Validation rules | Adaptive or event-based reminders |
| **Data Analysis** | Human coding<br>Quote retrieval<br>Memo writing | List/regex scripts coding<br>Inter-coder reliability tests<br>Pattern examination/analysis<br>Visualizing patterns<br>Counterfactual checks<br>Data linking<br>Network overlays<br>Network simulations | LLM assisted coding<br>LLM assisted memos<br>ML classifiers<br>ML embeddings<br>Augmented Retrieval<br>Semantic Q&A<br>Typology validation |
| **Writing & Presentation** | Triangulation<br>Consistency checks<br>Real-time writing collab | Retrieval of analytic products<br>Generating visuals<br>Citation formatting/insertion<br>Plain-language summaries<br>Accessibility audits<br>Embedded runnable notebooks | Assisted writing<br>Assisted editing |
| **Sharing & Preservation** | Replication code<br>Notebooks<br>Codebooks<br>DOI archiving<br>Long-term preservation | Containerized analytic spaces<br>Interactive data portals/APIs<br>Tiered access controls<br>Encryption for sensitive data | Simulated participants |

Inset 2 describes common research stages and uses of computation at each.

### Workflow Stages and Computation

**Research Design:** At the outset of projects, computational tools are increasingly used in designing studies, reviewing literature, editing grants and IRB applications, and testing instruments like interview guides (Farber et al. 2025; Parker et al. 2023).
**Data Collection:** New applications extend computation into the field itself, with automated collection of interviews by chatbots, platforms for synchronizing fieldnotes, and reminders for structured data like time diaries (Astrupgaard et al. 2024; Chopra and Haaland 2023).
**Data Processing:** Once collected, qualitative data increasingly undergo systematic preprocessing with computational assistance, to deidentify and index information, particularly for projects conducted at scale (Li and Abramson 2025).
**Data Analysis:** The greatest proliferation of uses lies in analysis, which includes both traditional approaches that are streamlined and multi-method approaches combining computational modeling alongside in-depth interpretation of large or small scale qualitative data (Abramson et al. 2018; Nelson 2020; Pardo-Guerra and Phawa 2022).
**Writing and Presentation:** AI and automation support editing and even text writing or generating text as 'synthetic' participants (Kozlowski and Evans 2025) as well as allowing pattern visualization (Abramson et al. 2024; Hanson and Theis 2024).
**Sharing and Preservation:** The expansion of infrastructure enables the technical possibility for making qualitative findings more shareable (Abramson and Dohan 2015; Edin et al. 2025; Murphy et al. 2021; Freese et al. 2022).

The appendix contains a more detailed discussion with added references and debates. Our aim here is not to judge which applications are useful or correct, but to note how such tools have entered our repertoire with purpose, as well as drift. We now turn to an example of our own purposive application to workflow.

## Workflow Example: Team Based Ethnography, Supported by Computational Social Science (CSS)

This section outlines our use of computational tools in DISCERN—a five-year NIH-funded study (PI: Dohan) examining how older adults with dementia and their care partners navigate cultural, institutional, and structural conditions across three U.S. states. Data include 119 in-depth interviews, over 300 day-long site visits, and longitudinal immersion by multiple fieldworkers, building on protocols from prior team-based research (Abramson and Dohan 2015).

What we've called computational ethnography (Abramson et al. 2018) developed from a practical need to scale accurate, reproducible analysis of patterns in qualitative data on health, culture, and inequality. Our approach integrates computational tools with interpretive methods in iterative and transparent ways—supporting data organization, indexing, pattern recognition, and cross-checking in-depth data generated with people in real world settings. DISCERN is the focal case, but aspects of this workflow have been adapted by researchers working in other paradigms and settings (Li and Abramson 2025). We present the workflow in chronological order, highlighting how each step contributed substantively to the analysis. Rather than offering step-by-step technical instructions (included in the Appendix and GitHub), we focus on the value added relative to traditional use of QDA software, in terms of clarity, scale, and analytic insight.

**Research Design**

The study examined how people living with dementia (PLWD) and their care partners understand and navigate cognitive decline, with attention to variation across backgrounds and contexts among the more than 20 million U.S. adults directly affected by Alzheimer's disease and related dementias (ADRD). Site selection was informed by pilot observations, demographic data, and prior findings to maximize heterogeneity and include both confirmatory and disconfirming cases (Small and Calarco 2022; Bernstein and Dohan 2020). Project infrastructure used secure platforms (REDCap for regulatory data, Box for text file storage, an internal wiki-document for documentation) and version control (Git for code), integrated by weekly team meetings and a centralized, transparent system for protocols to coordinate work.

This study began pre-COVID and pre-GPT; no large language models (LLMs) were used in design. Literature reviews relied on standard tools like Google Scholar, libraries, and bibliographic software. All fieldworkers were fluent in the languages of their sites, trained in qualitative interviewing and field observation, and completed structured simulations with standardized patients (medical actors) to prepare for nuanced, respectful interactions with PLWD and care partners.

**Data Collection**

Field observations produced ethnographic fieldnotes that represented observed interactions. Interviews were conducted by trained researchers, at a location of interviewees choosing, and audio recorded. Transcription was completed by cleared professionals rather than automated services—costly, but precise and secure. Substantive identifiers (e.g., OA017 for older adult 17; MP002 for medical provider 2) enabled linkage across each paragraph of interviews, fieldnotes, and survey metadata, with pseudonyms substituted prior to publication. This system enabled easy data auditing, simple ways to generate counts of core respondent categories, easy organization in QDA software as well as computational tools, and the ability to examine change over time by organizing file naming to correspond to observation data. No data were uploaded to commercial AI systems or used to train commercial models. [6]

[6] A standardized file-naming convention linked site, date (in sortable YYYYMMDD format), data type (e.g., fn for fieldnote; iv1/iv2/iv3 for interview waves), participant ID, and researcher initials—enabling integration across systems. For example, CAsite1_20220908_iv1_OA013_XX indicates a baseline interview (iv1) with participant OA013 at a California site, conducted by interviewer XX. This format supported sorting by site or date, facilitated auditing and subsetting in QDA software and Python/R, and

### Data Processing

To prepare transcripts and fieldnotes for analysis, we cleaned and standardized files to ensure compatibility across tools.[7] This made it possible to move seamlessly between ATLAS.ti, Python-based scripts, and archives without changing the original text. We created a shared list of key terms—focused on dementia-related topics like symptoms, care settings, and social ties—to support automated tagging. These lists were clearly marked to signal when coding was machine-assisted and always used alongside close human reading. Scripts helped check formatting, track participant IDs, and confirm de-identification, with manual review at every step. This allowed both shared topics and flexibility for the team, as we were able to link each paragraph to both individuals and their metadata— e.g., site, demographics, and other self reports. It also improved our ability to compare or focus on particular categories (e.g. those navigating brain disease and poverty) and saved months of time, which meant the team could focus on collecting and memoing data rather than indexing identifiable attributes by hand (Li and Abramson 2025).[8]

### Data Analysis

Our analysis followed a broadly realist approach, using computational tools to extend the kind of iterative work common in qualitative research—guided by previous studies and reflecting patterns observed during time spent with people: listening, observing, and engaging in everyday life (Abramson and Sánchez-Jankowski 2020).

We treated "coding" as a layered process for tagging segments of text so they can be easily found, re-read, and analyzed. Our goal was to build an index of meaningful concepts—not to replace deep interpretation. We use the term *indexing* to emphasize this technical function and to avoid the assumptions that coding always reflects a final interpretation or full theory. The length of text units being tagged should reflect analytical goals. We indexed interviews and fieldnotes at the paragraph level.[9]

We used four main styles of indexing across projects, with CAQDAS (ATLAS.ti) and python:

---

linked data to REDCap quantitative measures (Li and Abramson 2025). Adjacent digital instruments (e.g., diaries, geolocation) were possible but not deployed. Consent records, site logs, and tracking sheets were stored separately in secure systems, with downloadable metadata linked through REDCap.

[7] Files were reviewed in Sublime Text to remove stray characters, enforce UTF-8 encoding, and standardize breaks and whitespace. This was verified in a python pre-processing script. Data were organized for interoperability across ATLAS.ti, Python, and archival systems. Keyword matching used grep/regex and was flagged with an auto_ prefix to denote potential false positives. Embeddings (BERT, RoBERTa, Gemma2-2B) supported synonym expansion, with all steps linked to original text for verification.

[8] We made light adjustments to text (like consistent punctuation and lowercasing) to support computational tools while preserving the original language for interpretation. Language models helped identify empirically related terms through semantic similarity metrics, and fieldworkers contributed throughout—ensuring that the tools reflected grounded knowledge. This kind of hybrid approach helps researchers handle more data than can reasonably be coded by hand (Abramson and Dohan 2015; Li et al. 2021). Current iterations of work may add deidentification pipelines, to auto replace names or places, e.g. Oakland→ [[city1] (DuBois et al. 2023).

[9] This is common in medium- to large-scale work that involves primary qualitative data sources like fieldnotes and interviews (e.g. greater than 500 pages of text). Paragraph-level indexing is a pragmatic middle ground: it is precise enough to tag specific ideas, preserves context, supports overlapping themes that smaller units miss, and returns a manageable amount of text in software queries. Full documents are often too broad a unit for indexing interviews or fieldnotes, even though document-level coding is well established and useful for some large-scale document analysis. For example, if social networks and health challenges are both on the interview guide, they may appear in every transcript—making document-level co-occurrence uninformative. Paragraph-level indexing shows when and how these concepts actually appear together in the same narrative context. Sentence-level tagging avoids false positives but can miss overlaps that unfold across longer spans (e.g., separated by a sentence) (Abramson 2011; Li, Dohan, and Abramson 2021).

1. **List/Dictionary Indexing**: For high-certainty concepts (e.g., clinical trial mentions), we used rule-based searches with regular expressions (regex)—a compact syntax for finding patterns in text, supported by both commercial and open-source tools. [10] This allows efficient tagging across large datasets. We applied this method to flag common topics, link data by ID codes (e.g., DR001 for each paragraph of interview speech and observed interactions with a doctor), and segment longer texts into paragraph units for further indexing (Abramson and Dohan 2015).
2. **Full Human Coding**: For nuanced, emergent, or interpretive concepts (e.g., identity), we relied on traditional human coding rooted in close reading and memoing. This remains critical for meaning making in complex or sensitive domains and can also be used to inform lists (as in #1) or train local models (as in #3).
3. **Hybrid Machine Learning (ML) Classification**: To scale moderately interpretive tags (e.g., social networks, place references), we fine-tuned local ML classifiers using examples from human-coded data and then used them to index broader corpora for patterns (e.g., identifying barriers_to_healthcare). All machine-generated labels can be reviewed and refined by researchers to ensure conceptual accuracy (see Li et al. 2021).
4. **Pattern Discovery and Visualization**: For exploratory analysis or mapping relationships, we used tools that go beyond traditional codes. Word embeddings—numerical representations of text meaning—identify concepts used in similar ways (e.g., dementia_mention and framing_aging_as_negative). Semantic networks map how concepts co-occur in text, such as how people discuss success or describe pain in moral terms that vary by group (Abramson et al. 2024). These tools can surface patterns that merit further attention or help confirm expected relationships (Abramson et al. 2018) but are always interpreted alongside close reading to avoid conflating syntax with substantive meaning (Boutyline and Arseniev-Koehler 2024).

This layered approach supports a range of analytic strategies for theoretical works, applied policy research, and sharing with communities.
Indexing types served complementary roles: list/dictionary tagging provided broad structure; human coding grounded interpretation; classification scaled complex patterns; and pattern discovery (see Table 3). In some projects, we scaled human-coded categories using local, small language models run securely on in-house servers; these extended codes across the dataset in under one-third the time of manual coding, with secondary human review ensuring interpretive accuracy (Li et al. 2021). [11] Each paragraph remained linked to its full document source using positional metadata in both the QDA software and our Python schema (see Appendix).Weekly meetings integrated field and analytic work: fieldworkers shared observations; analysts presented outputs; memos documented interpretations, disconfirming evidence, and evolving insights. We reflected on reflexivity, site variation, and community feedback. Identifiable outputs were restricted internally; public materials were de-identified. Automation was used when consistent with our methodological and ethical commitments. No data were shared with commercial large

---

[10] Example of a regex pattern used to identify variants of "clinical trials" in text: "\b(clinical\s+(trial(s)?|study|studies|research|experiment(s)?)|randomi[sz]ed\s+(controlled\s+)?trial(s)?|RCTs?|first[-\s]in[-\s]human\s+(trial(s)?|study|studies)|phase\s+(I{1,3}|IV)\s+trial(s)?)\b ". This expression matches commonly used phrases including *clinical trial(s)*, *clinical study/studies*, *clinical research*, *clinical experiments*, *RCT(s)*, *randomized controlled trial(s)* (both US/UK spelling), *first-in-human* (hyphenated or not), and *phase I–IV trials*. The \b metacharacter denotes word boundaries, preventing partial matches (e.g., avoiding "nonclinical"), and the | operator means "or"—allowing multiple variants to be matched within a single pattern. Current QDA platforms and languages like Python allow integration into automation and searches.

[11] We used an F1 score of 0.80 to assess human–machine coding agreement, and a Krippendorff's alpha of 0.80 for intercoder (human–human) reliability. These tests use established open-source models like BERT variants (Li and Abramson 2025).

language models; local models were used strictly for scaling or surfacing patterns for human review.[12]

**Table 3: Qualitative Indexing Approaches and Uses***

| | When to Use | Purpose/Strength | Cautions/Limits |
|---|---|---|---|
| **List / Dictionary Indexing** | When tagging well-defined, high-certainty concepts with clear linguistic boundaries (e.g., "clinical trials," "transportation_mention"). | Enables rapid, consistent tagging across large volumes of text; supports fully reproducible search and comparison. | Risks missing nuance or context; false positives; should be limited to terms with stable meaning. Use prefix in QDA software (e.g., auto_symptom) to flag for review. |
| **Full Human Coding** | When analyzing interpretive, context-dependent, or emergent categories (e.g., age_stratification, illness_stigma). | Preserves contextual richness and interpretive depth; essential for meaning-centered or theory-building work. | Time-intensive; inter-coder reliability and review needed to ensure consistency. |
| **Hybrid: Human-Machine Learning** | When extending researcher-defined codes (e.g., social_network_mentions, health_disparity) across very large corpora. | Scales human classifications through supervised learning; combines automation with iterative review; reduces repetitive tagging. | Requires quality human training data and setup; models can drift or amplify bias if not used well; secondary human review essential. |
| **Pattern Discovery / Modeling** | When exploring co-occurrence, conceptual relationships, or latent meaning across and within text data (e.g., word embeddings, semantic networks, topic modeling). | Reveals associations and structures not immediately recognized; useful for visualizations and confirming patterns. | Results are probabilistic and must be interpreted in dialog with qualitative data; always verify patterns by returning to source text. |

**These can be combined in a project, such as using lists to index basic terms or segments, machine learning to tag more complex topics, and human coding to classify deeper distinctions.*

## Writing and Dissemination

Text analysis and ethnographic interpretation were mutually informing, a key component of computational ethnography. Heatmaps summarized case-by-code patterns in analyses of words and cases (Abramson and Dohan 2015; Arteaga et al.2025); semantic networks provided meso-level maps linked to in-depth narratives; targeted retrieval grounded claims in verbatim passages and linked back to the original. Findings were reported as patterns across interviews and observations, but we also focused on exemplary and disconfirming cases when relevant to research questions. Pseudonyms and de-identification were used to protect participants and meet IRB requirements. AI was not used in writing, though local models checked for typos on deidentified transcripts. Community partners engaged with visuals and findings as in earlier projects.

[12] Language models used here were trained to identify patterns, not generate text, and were implemented locally to maintain data security and control. Paragraph-level links were maintained using positional metadata compatible with both QDA and Python systems, as described in the schema in the appendix.

**Reflection**

The workflow described above was developed to scale and streamline team-based qualitative research, though we have also applied it in individual projects. This process has produced more accurate coding in all of our tests, supported making our data usable for multiple projects and audiences, helped identify patterns missed in human review, and enabled systematic comparisons at various units of analysis— people, field-sites, documents, narratives, and text segments. The necessity of varied computational infrastructure for this and other large projects extends beyond analytical preference because traditional CAQDAS often struggle or break, necessitating workarounds and additional tools. However, even in our DISCERN example, computational tools extended rather than replaced qualitative insights.

Coordinated practices—such as meetings, file-naming, analytic schemas, automation, and human reading—played a central role in integrating technical and human aspects. We used them to collectively and individually respond to broader methodological challenges, including framing, reflexive understanding of field positions, variations in what we saw; decompress from stressful encounters; identify what constitutes a case; and discuss which comparisons are valid or unreasonable (Abramson and Gong 2020). We treated these as epistemic and ontological choices that involved but were not dictated by technological tools (Cicourel 1964; Dohan and Sanchez-Jankowski 1998; Pardo-Guerra and Pawa 2022). At the core, indexing while retaining access to full text allows for re-reading and supports a range of outputs—from academic writing to policy briefs and public commentary—which is core to in-depth qualitative work.

This is not a prescriptive model, but a pragmatic illustration: researchers can adapt or reject components based on philosophical commitments and project needs. Variation and disagreement each can help clarify limits, model alternatives and improve methodological practice.

**Pragmatic Workflow Ideas for Solo and Smaller-Scale Research**

These principles apply to solo researchers and lower-resourced projects, including comparative ethnography using student licenses, text editors, and lightweight CAQDAS (cf. Abramson 2015; Dohan 2003).

**Prioritize Open-Source Tools**. Especially when working with language models, consider open-source options that can run securely on local machines. This avoids privacy risks and commercial costs while providing flexible capacity for text analysis using tools like Python or R.
**Build a Scalable Data Structure**. Consistent file naming and clean formatting from the start make your data easy to track and machine-readable. Simple structures help maintain order as your project grows—with or without automation.
**Automate Strategically**. Use automation where concepts are clear and repeatable—like dictionary or regex-based indexing—but reserve deep coding for interpretive themes. Don't reduce explanation to indexing systems. Memos and human reading remain central.

**Centralize with QDA or Markdown Tools**. Tools like ATLAS.ti, MAXQDA, or markdown editors (e.g., Obsidian) help link raw text, memos, and outputs from scripts. Even if not coding with scripts, structured naming, memo consistency, and clear documentation support analytic coherence. Always back up data securely.

**Memo Frequently**. Memos bridge data and writing. They matter regardless of coding system or epistemology—and are just as useful in programming or policy work as in qualitative interpretation.

**Know the Limits**. For small datasets (e.g., <20 interviews), automation, and even CAQDAS, may not add value for some approaches and can divert attention. Yet, tools like heatmaps remain useful for small-scale qualitative projects (Garrett et al. 2019; Mische and Mart 2025), as well as large-scale quantitative and qualitative analyses (Abramson et al. 2018; Fosse 2023).

**Learn Through Application**. Focus on learning what serves your current project. This makes complex tools approachable and connects skills directly to research needs.

# Discussion

Qualitative work and computation have long been intertwined, from early formal modeling and qualitative data analysis (Dohan and Sánchez-Jankowskii 1998; Mohr and Duquenne 1997) to narratives (Franzosi 1998), digital repositories (Murphy et al. 2021), and recent interpretive-computational projects (Abramson et al. 2024; Nelson 2020) as well as quantified text (Roberts et al. 2022). We have discussed enduring concerns (e.g., whether software imposes a singular logic), technical possibilities (e.g., scaling and visualization), and pitfalls (e.g., environmental and scientific costs) in an era of AI.

Though predicting futures is fraught, one projection seems clear: whether computation is cast as a pinnacle of human achievement, a mechanized cage of algorithms, or a toolkit with potential to harm or help, it is unlikely to disappear so long as society continues in its current form. It saturates everyday life, the worlds we study, and our research, even if we might hope otherwise. How we respond is a matter of bounded agency. Responses will vary even within qualitative fields—a strength, perhaps (Abramson & Gong 2020).Our bias is toward pragmatic and purposeful engagement with tools, which, if used well, can yield accurate insights into health, inequality, culture, labor, and science, even as they carry potential for harm—much like research itself (Abramson 2024).

*Keeping up with the Algo-Joneses*

In discussing the challenges of technology, one theme that comes up but is rarely mentioned is the frenetic pace of development. The phrase *keeping up with the Joneses* refers to the pressure to match others, often in a competitive way. This remains a very real concern in an era of not just computation but quantified scholarship, constantly changing models, and the moral valence of either embracing or rejecting algorithmic tools. Hence, keeping up with the "algo-Joneses" (algo signifying algorithm). Classical sociologists noted that such dynamics can become endemic to modernity—producing systemic inequality and driving individuals to pursue unattainable goals while losing sight of broader social forces (Durkheim 1997; Weber 1904). Contemporary studies of digital society extend these observations, documenting how rankings, feeds, metrification, and digital traces affect life chances (Fourcade & Healy 2024; Mears 2023; Pardo-Guerra 2022).[13]

Keeping up with the "algo-Joneses" can feel futile and mandatory. Yet, as we have shown, possibilities for thoughtful engagement exist. The point is neither to weigh down scholars with despair nor to offer hollow optimism, but to recognize the coexistence of contradictory currents—itself a central feature of sociology. Technology has long been both a tool of creativity and an avenue of control. The question is how to respond when the algo-Joneses continue to bring home new tools, some useful, others alarming.

*Considerations from the Present and Future*

We offer the following suggestions from our use of, teaching about, and study of technology in qualitative research:

1. **Computational literacy matters.** Just as quantitative and qualitative skills are both vital in a multi-method field like sociology, awareness of computational tools is part of engaging with the present and future of social inquiry. Algorithms, text analysis, AI and even simulations using text data are increasingly widespread. Even those who resist this computational expansion will benefit from understanding systems in order to engage critically. This is challenging—some tools are opaque and the field is evolving—but important. While the ability to work by hand may remain essential, doing only that may increasingly become untenable for some projects.
2. **Local systems help mitigate downsides.** Running small language models on a desktop may lack the spectacle of GPT-style 'chatting' interfaces, but it enables practical tasks: fine-tuning transformer models to check codes, visualizing patterns, and validating analysis securely without uploading sensitive data. While all research tools are embedded in broader commodity chains, local use reduces environmental and confidentiality risks often assumed inherent in computational work.
3. **Human judgement remains central.** Automating key decisions risks undermining interpretation. At the same time, entirely manual workflows can exhaust capacity for higher-level reasoning. The boundary between automation and manual work is context-dependent but human oversight remains central to qualitative analysis even with computing (sometimes called "human in the loop" in computational settings).
4. **Disagreement is inevitable—but can be constructive.** Rejecting computation entirely—or dismissing those who question it—closes off dialogue. Whether you use QDA, STATA/R, Python, or none of the above, core goals of uncovering patterns and understanding are shared . One of the strongest contributions of fieldwork is providing perspective, and multiple perspectives offer possibilities for learning and improving in accordance with sociological pluralism—especially within big tent approaches like qualitative research (Abramson and Gong 2020; Lamont and Swidler 2014).

---

[13] For qualitative scholarship, the issue is close to home. The pressures to adopt new technologies can be understood through a scientific field shifting towards easily quantifiable outputs over long-term ethnographic work (Pardo-Guerra 2022). Google Scholar, for instance, systematically under-indexes book-to-book citations, devaluing the currency of ethnographic monographs while amplifying the metrics of shorter, digitized outputs. The effect, already in motion before the rise of generative AI, exacerbates pressure for everyone to produce more rather than better, and encourages conservative uses of technology for efficiency rather than improving inquiry. Those earlier in their careers, or in less secure environments, are in a more precarious position as the stakes involve the ability to continue to work in social science and earn a living with hard won skills (Fine and Abramson 2021).

5. **Sociology has a legacy of novel methods and triangulation.** From W. E. B. Du Bois's data visualizations at the 1900 Exposition to repurposing computing systems to chart inequality during COVID, sociologists have long combined data types and methods, repurposing technologies in the pursuit of understanding. The same potential exists for AI—at least in its broadest sense.
6. **Shared formats make work easier to reuse and build on.** A major barrier to replication, data sharing, and integrating computational tools is the lack of a simple, durable format for organizing data. Many file types—especially those tied to specific software—are overly complex or hard to work with outside that system. The field would benefit from using a straightforward format (like the example we share in the appendix) that works across both qualitative software and programming tools. It doesn't need to do everything—just enough to support communication, collaboration, and targeted checks in larger projects.
7. **Open, critical engagement is key.** Rather than rushing to keep up, scholars might adopt a stance of openness to learning about emergent technologies while critically assessing their limits and possibilities. What works in one case may not in another. Still, sharing tools and methods can strengthen collective practice, even if doing so is not labelled "open science."

*Concluding Thoughts*

We write in a moment of rapid transformation, when the intended and unintended uses of computation and AI demand attention to influences, risks, and possibilities. We have argued for a pragmatic approach to qualitative analysis and computational tools: use technology that can advance methodological principles, thoughtfully and defensibly. This is not an embrace of technology as progress, nor a rejection rooted in dystopian fears. Rather, it reflects the view that dualisms are a luxury. Today, every approach to social life entails some position on computation—explicit or implicit, deliberate or de facto. That is both methodologically consequential and unlikely to change.

To paraphrase Weber on rationalization and bureaucracy: the future is uncertain, but will likely include powerful technologies—including AI—that constrain and enable social life, in ways that aren't simply "click, to opt out." Our aim here has not been to decide whether those developments serve a greater good, but to connect divergent positions to real methodological questions, ethical concerns, and practical strategies for responsibly using computation in qualitative research.

We hope that by:

1. Clarifying intersections between qualitative research and computation;
2. Outlining a typology of engagement in an era of automation and AI;
3. Walking through existing workflow decisions; and
4. Reflecting on the challenges and possibilities of pragmatic sociological practice—

—we've offered not just ideas, but tools to support inquiry and human agency.

For added implementation details, references, and code, see the footnotes, appendix and linked repository.

**Appendix references**

# APPENDIX & SUPPLEMENTS

Resource Page on Computational Social Science for Qualitative Research (tool list, code repository, and versioning logs) is available at [GitHub link] version 2025-10-01.

- Computational Resources for Qualitative Resources: https://github.com/Computational-Ethnography-Lab/teaching/

# A. WORKFLOW

## I. TABLE 2 EXEGISIS

This section elaborates common phases of a project, and some potential uses of computation.

*Research Design.* At the outset of projects, computational tools are increasingly used in designing studies and instruments. For instance, in some studies large language models (LLMs) are used to synthesize literature for systematic reviews through both custom api and growing commercial systems (Parker et al. 2023). LLMs are also used to test the readability of interview questions prior to piloting, to help with initial translations on recruitment fliers, and to check for typos. Many of these options existed in some form in word processing software, but the tools have expanded and barriers to entry are lower. Some simulate potential participants or scenarios to stress-test protocols (Kapania et al. 2024; Kuipers 1986) and use in field experiments is expanding. Project management and version control systems (e.g., OSF, GitHub) support collaborative design and documentation, sometimes including automations but also usable without AI (Farber et al. 2025).

*Data Collection.* New applications extend into the field itself. Automated transcription services are replacing human transcription in many venues (Zhou 2025) and digital consent platforms may aid in the capture of field materials or digital documents—representing a potential way to connect surveys, time diaries, and qualitative accounts. Researchers have employed mobile platforms to collect diaries and ecological momentary assessments, augmented by automated reminders or AI-generated prompts (Farber et al. 2025). In a controversial data collection application, AI agents have been deployed as interviewers (Chopra & Haaland 2023). The argument is that chatbots can conduct structured text collection, sometimes enabling participants to feel more comfortable sharing sensitive experiences than with a human (Chopra and Haaland 2024). To our knowledge, this remains largely rhetorical or at least unmeasured: growing evidence indicates that the conventional claim—that in-person interviews yield more accurate data—is not consistently substantiated across settings, and the technology is changing quickly (Johnson et al. 2019).

*Data Processing.* Once collected, qualitative data increasingly undergo systematic preprocessing with computational assistance, particularly for projects conducted at scale. Through a combination of named entity recognition (NER) and human review of the AI processing on a local computer, current technologies can facilitate de-identification of transcripts to protect confidentiality. Scripts written in Python or R can automate the conversion of text, audio, and images into standardized formats for computational analysis and easier human reading, while metadata tagging and automated quality checks can enable analysis by data type, demographic group, or topic in qualitative data analysis software (Li and Abramson 2025). These steps extend earlier practices of structure and organization (Dohan and Sánchez-Jankowski 1998) into more formalized, computationally reproducible pipelines that can be shared as reproducible code in accordance with the principles of open science (Freese and Peterson 2017; Gupta et al. 2021).

*Data Analysis***.** The greatest proliferation of uses lies in analysis, at least by current estimates. For instance, hybrid approaches to coding integrate machine learning classifiers or LLM-assisted coding into qualitative research. Uses range from largely inductive (searching for emerging patterns) to deductive (applying existing categories), in ways that reflect various approaches to social science. Computational grounded theory (Nelson 2020; Alqazlan et al. 2025) exemplifies an approach in which algorithmic analyses help surface patterns inductively. Analysis is refined through human interpretive reading to parallel a focus on emergent themes. In another approach, expanded iterative qualitative coding, iterative coding (Deterding and Waters 2021) – already augmented with basic automation in flexible coding – has been expanded to include hybrid pipelines that allow language models to use human codings to learn how to classify text. Human researchers review the results and conduct higher-order interpretation, using both deductive concepts (e.g. study domains or hypotheses) and inductive findings (e.g. emergent patterns), as part of the indexing process (Li et al. 2021). This can be done with high consistency and accuracy using machine learning on secure local computers, sometimes identifying relevant data humans may have missed. For example, Yoon and McCumber (2024) used a combination of natural language processing methods to inductively construct a topic-by-geography typology of New York Times Travel articles, which then guides their qualitative analysis of the symbolic hierarchy in the cultural narratives about places. Finally, the deductive application of categories, once a hallmark of quantitative content analysis and some formal approaches can be applied at scale with good accuracy using a variety of models (Than et al. 2025).

*Data generation and simulation.* With the rise of large language models, some researchers have begun experimenting with using large language models to generate synthetic data in the social sciences. A common design is to prompt models to simulate "empirically realistic, culturally situated human subjects" or "personas" by reproducing real survey responses from the General Social Survey or World Values Survey (Bisbee et al. 2024; Boelaert et al. 2025; Kozlowski and Evans 2025). These outputs, however, tend to be more biased, less variable, change over time, and oversimplify identity groups compared to real human responses (Argyle et al. 2023; Bisbee et al. 2024; Boelaert et al. 2025; Kozlowski and Evans 2025; Wang, Morgenstern, and Dickerson 2024; Zhang, Xu, and Alvero 2025). Commercial models are also shaped by "model alignment"—modifications by companies to make outputs match user expectations or avoid controversial content (Ji et al. 2023). While alignment may reduce harmful language, it can also limit usefulness for studying prejudice, misinformation, or other social processes (Lyman et al. 2025).

Even researchers who don't use AI to generate data face these issues during analysis. People increasingly use LLMs to answer open-ended survey questions (Zhang et al. 2025), bots account for an estimated 20% of social media interactions, and many now use LLMs—introducing AI-generated text into research materials (Ng and Carley 2025; Radivojevic, Clark, and Brenner 2024). While some organizations resist LLMs for data security (Serbu 2024), others—including universities—embrace them (Singer 2025), accelerating synthetic text in administrative records, archives, and qualitative data. As AI-generated content spreads, researchers may need to account for it throughout qualitative data analysis, not just in study design.

*Writing and Presentation.* AI and automation support writing and dissemination, though both are points of contention. LLMs (both offline and commercial) can draft memos, generate figures, summarize and can create tables from coded or uncoded data (Farber, Abramson, and Reich 2025). Collaborative platforms allow for live editing, and literature agents automate the ability to web scrape scholarly articles and link to bibliographic databases. Word processors have long included spell checks, but increasingly use advanced tools or apps (e.g. Grammarly) to edit, rewrite, and even write academic work, raising possibilities for clearer writing, as well as concerns about subtle adjustments to scholarly positions (and fabrication). In terms of presentation, computational tools allow for novel visualizations of qualitative data from ethnoarray heatmaps (color coded tables of people, sites or interaction) that show patterns of cases (Abramson and Dohan 2015) to embeddings and network analyses (Hanson and Theis 2024; Basov et al. 2020). A rich tradition in cultural sociology has been using network, algebraic and vector structures to visualize spaces of cultural meanings (Bearman and Stovel 2000; Breiger 2000; Carley 1994; Mohr 1994). Recently, using word embeddings (vector representations of language) to map culture in historical texts and subject speech as part of broader examinations of patterns of meaning, is gaining traction in a variety of fields (Boutyline and Arseniev-Koehler 2025; Kozlowski et al. 2019).

*Sharing and Preservation:* The expansion of infrastructure enables the technical possibility for making qualitative findings more shareable, with code and analytic notebooks enabling replication—or websites linking text segments for interaction– though questions of understanding, representation and interpretation remain central to sociological approaches (Abramson and Dohan 2015; Murphy et al. 2021). Researchers are also using AI to document their work and workflow in new ways. These include analytic processes like "promptbooks" (Stuhler et al. 2025) and extending the idea of a living codebook that captures indexing procedures in the era of AI through charting the evolution of computer instructions to index themes and patterns (Reyes et al. 2024).

Taken together, the applications above highlight computation-enabled workflow changes and possibilities, ranging from incremental to potentially transformative. Some are common (e.g., spell checking, using Google Scholar for literature), others specialized or emergent. Many practices using computation remain recognizably qualitative: iterative coding, interpretive reading, theorizing from cases, combining induction and deduction to speak to explanations, visualizing data, and triangulating analyses have deep roots in sociology. Yet each stage of the research process is increasingly mediated by computational tools that promise efficiency, scalability, and transparency—while raising enduring concerns about validity, bias, and epistemological fit (Ashwin, Chhabra, and Rao 2023; Amirova et al. 2024)—whether their use is purposeful or simply determined by the technology itself. For example, even basic web use, of systems like Google Search, are mediated by algorithms in ways beyond user control (Noble 2018; Pasquale 2016).

## II. CHECKLIST AND CONSIDERATIONS

**Checklist: Solo/Small-Scale Computational Qualitative Workflow**

1. Clarify your goals and whether computation will help.
2. Use accurate transcripts—manual or reviewed automated services.
3. Use consistent file names that include site, date, ID, and interviewer.
4. Save files in UTF-8 format for software compatibility.*
5. Use regex to search key terms or entities.
6. Build a transparent keyword dictionary (e.g., terms related to dementia).
7. Track codes with spreadsheets or lightweight QDA tools.
8. Only use embeddings locally, or with de-identified text.
9. Keep original paragraphs and metadata for clarity.
10. Track all edits and versions (e.g., using Git).

*See footnotes, there are free or open sources tools for UTF-8 encoding and tools like Sublime.

**Table AI: Risks and Considerations  for Computational-Qualitative Work**

| Risk | Why It Matters | Mitigation |
|---|---|---|
| Uploading raw transcripts to commercial AI tools | May breach confidentiality agreements or institutional review board terms | Use anonymized data and secure systems (e.g., local LLMs, private APIs) |
| Relying only on dictionary or keyword methods | Can miss context and nuance; reduces complex meaning to word counts | Combine automated tagging with close reading and interpretive coding to ensure match |
| Ignoring version control | Makes tracking changes and team collaboration difficult; undermines transparency | Use Git or cloud-based systems to document all changes with timestamps |
| Vague or missing data documentation | Creates confusion about what data represents; prevents others from understanding or reusing materials | Create structured documentation specifying interview conditions, participant characteristics, and coding decisions |
| Assuming saturation early | Stops data collection before capturing full range of experiences or perspectives, disconfirming cases | Document when new themes emerge; actively seek cases that challenge initial patterns; justify sample decisions explicitly |
| Not reviewing automated annotations | Algorithms make systematic errors that go undetected | Check model outputs against human interpretation (and perhaps, vis versa); document agreement levels and error patterns |
| Analyzing excerpts without surrounding context | Loses meaning shaped by conversational flow and speaker relationships | Preserve links between coded segments and full transcripts; note what comes before and after |
| Treating model outputs as neutral or objective | Models embed specific assumptions and biases in their design and training | Question where automated coding succeeds and fails; document analytical choices behind computational approaches |
| Ignoring biases in model training data | Models reproduce worldviews of their training materials; systematically misinterpret underrepresented groups | Assess whether training data reflects participant experiences; acknowledge when models may distort marginalized perspectives |
| Treating interviews, documents, and observations as interchangeable | Different materials require different analytical approaches; conflating them obscures their distinct properties | Develop analysis strategies appropriate to each data source; document what each type of material can and cannot reveal |
| Letting computational capabilities determine research focus | Pursues questions because they are technically feasible rather than theoretically important | Define substantive questions first; select or advance methods that address those questions rather than reshaping inquiry around available tools |

## III. RESEARCH TASKS AND USEFUL TOOLS (AS OF OCTOBER 2025)

We did not necessarily use all tools listed above in our projects, nor is this an endorsement. The examples above illustrate how computers plug into contemporary workflows. Further, even basic word-processing, spreadsheets and code platforms now include AI features and may generate data about use, possibly unbeknownst to user—configure settings accordingly.

### Table A2: Examples Computational Tools and Uses by Research Stage

| Research Stage & Need | Example Tools | Example Use(s) |
|---|---|---|
| ***Design*** | | |
| Literature review & synthesis | * Elicit; * Consensus; ‡ GPT (with web search); ‡ Gemini Deep Research | Triage and summarize relevant articles; compare findings; export key notes to your repo or Zotero. |
| Planning & team coordination | * Slack; * Asana | Slack channel for daily field logistics; Asana task with deadlines to aggregate work. |
| Secure storage & access | * Box; * Dropbox; * Data enclave (institutional) | Store data and audio in the data enclave; share de-identified working materials via Box/Dropbox with appropriate access controls. |
| ***Data Collection*** | | |
| Participant tracking & consent | * REDCap; * Qualtrics | Use an access-controlled tracking database (REDCap or Qualtrics) for consent, contact details, and recruitment status. |
| Fieldnotes | * Obsidian; * NotePlan; *Sublime Text | Keep fieldnotes on Box or within the app; type quick notes during/after visits in Sublime Text. |
| Audio capture & transcription | * Digital recorders; * Rev; * Otter.ai; † Whisper | Record on a digital recorder (with a second as backup); upload to an approved program with NDAs/insurances; opt out of model training if using AI services unless consented. |
| ***Data Processing & Analysis*** | | |
| De-identification (text) | † spaCy (including medspaCy); † scrubadub | Natural language pipelines in Python: run spaCy/medspaCy NER to flag entities; use scrubadub for simple redaction; human review confirms edits before export. |
| Indexing / qualitative coding (paragraph-level) | * ATLAS.ti; * NVivo; * Dedoose; * MAXQDA | Indexing pass (paragraph level), list-based tagging, then in-depth coding; automate where useful while considering whether data should stay local. |
| Patterned tagging | † Python re (regular expressions) | Use the "clinical trials" regex to flag matching paragraphs and export matches with 1–2 paragraphs of surrounding context for review. |
| Quant text analysis (R) | † quanteda; † stm | Build a document-feature matrix in quanteda; fit stm with covariates (e.g., site/wave); optionally visualize with a simple heatmap by group. |
| Quant text analysis (Python) | † pandas; † scikit-learn; † BERTopic; † spaCy | Tokenize/NER with spaCy; vectorize and cluster with scikit-learn; use BERTopic for topic overviews; optionally heatmap feature/cluster co-occurrence. |
| Classification & summarization | * Ollama | Run a local transformer model for text classification or a small generative language model to summarize a single transcript on-device. |
| Visualization (heatmaps & comparisons) | † R base heatmap; † matplotlib, †seaborn, † plotly | Create a basic heatmap to compare codes/topics across sites/waves and surface contrasts for memoing. |
| ***Writing & Presentation*** | | |
| Writing & code review | ‡ Claude; ‡ GPT; ‡ Gemini (API modes) | Use for writing clarity and code-comment review with version/temperature controls; keep claims tied to source paragraphs. |
| Writing & compilation | * Microsoft Word; * Overleaf; * Pandoc; * Grammarly | Draft in Word; use Overleaf for equations/tables when needed; Pandoc converts analysis outputs; Grammarly for readability checks. |
| Citation management | * Zotero *Mendeley | Collect citations/notes; insert references into Word/Overleaf; organize PDFs and annotations. |
| ***Sharing & Preservation*** | | |
| Reproducibility & teaching repo | * GitHub; * OSF; * Jupyter Notebooks (Markdown); * Quarto | Version analysis notebooks/scripts on GitHub; mirror essential artifacts on OSF; keep narrative in Markdown cells; render reports with Quarto. |

Legend: * Program/App; † Package/Library; ‡ LLM/Model.

# B. PROCESS AND OUTPUT EXAMPLES

## TEXT FORMATTING

Output is ideally deidentified, and checked for proper encoding, formatting for machine readability and interoperability between a data frame and QDA software whenever shared beyond close collaboration of a core team. De-identification involves the replacement of terms like Houston with [city], at a level determined by analysts, laws, or policy. This may range from HIPAA level redaction for medical records, or simpler compressions for typical social science data.

Even if data is secure and not de-identified, formatting text (as well as including IDs, like CG### for caregivers) has implications for linking data, search, retrieval, and computational analysis.

*Interview Formatting*

Our standard is two hard returns between interviewer and interviewee. One hard return between paragraphs on multi-paragraph answers. Always using ID: to indicate speaker, and allowing us to subtract interviewer speech from modeling  with natural language processing (Li and Abramson 2025).

> INTV_BG:     Lovely. Do you know if she's in the memory care area, or if she is assisted living, separate from?
>
> CG021:  She is technically in memory care, but she is possibly the person that needs the least care, from my observation, on her floor. Her floor is where other people come to have a meal, or the lounge, where they do the games, and the harp. I mean, it's really nice. There's a harpist that comes. It's really fancy <laughs>. Although, though, it is called memory care, yeah, because she has to have some help with, certainly, taking her medicines, and just nudging her to having good, healthy behavior. But I think they help her with laundry. There are some things that she was still doing at home. Like she could make her own tea and coffee, and heat up food. But she wouldn't really eat very well. Even with really healthy options, she would still want to just eat ice cream, root beer floats <laughs>.

*Fieldnote formatting*

Use paragraphs for interactional sequences. Use "" for verbatim quotes. Indicate your field position. Use [[]] to indicate personal experiences or commentary. Use !!! for critical interactions. Use IDs for subjects, like OA003. Python script may adds Z_FN (if no unique id) or IDS of subjects for data linking in QDA software and outside.

> Z_FN: I decided to approach the ladies at like 12:30pm to ask if they knew if the center was opened or when it would open. I approached them and in Spanish asked, "Disculpen, Saben si el centro esta abierto" – "Excuse me, Do you know if the center is open" they then continued to respond in Spanish. They said that it was open but that the staff was in

*Making text machine readable*

Our computational data_frame uses a unified schema, to make text machine readable, post-cleaning, and connected to metadata, often using tabulated QDA export. See Li, Dohan and Abramson 2021 and example of subsetting [here](). Only text and document are required fields from the code examples we give, and others can be derived from file names or related data bases (like data_group".

```
# Schema with Python typing
schema = {
    "project": str, # List project, e.g. DISCERN
    "number": str, # Position information, e.g. chronological ID
    "reference": int, # position information, e.g. added information
    "text": str,  # Content, critical field: must not be empty, e.g. actual text
    "document": str,  # Data source, Critical field; e.g. OA010_iv1_20250411.txt
    "old_codes": list[str],  # Optional: codings, must be a list of strings, for codes not active
    "start_position": int, # Position information; character and QDA info
    "end_position": int, # Position information; character and QDA info
    "data_group": list[str],  # Optional, to differentiate document sets: Must be a list of strings
    "text_length": int, # Optional: NLP info
    "word_count": int, # Optional: NLP info
    "doc_id": str, # Optional: NLP info, unique paragraph level identifier
    "codes": list[str]  # critical for analyses with codes, Must be a list of strings; active filter codes
```

}

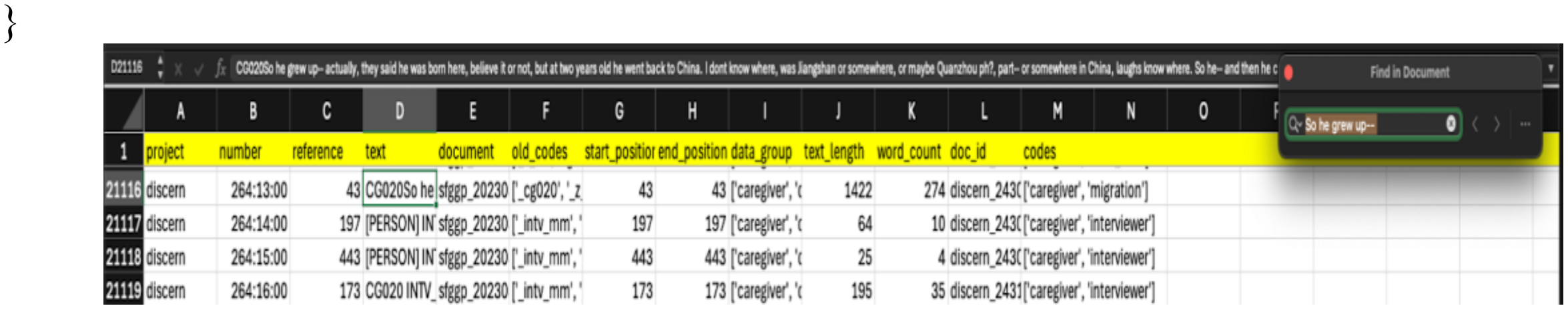

D21116 CG020So he grew up-- actually, they said he was born here, believe it or not, but at two years old he went back to China. I dont know where, was Jiangshan or somewhere, or maybe Quanzhou ph?, part-- or somewhere in China, laughs know where. So he-- and then he c

| | A | B | C | D | E | F | G | H | I | J | K | L | M |
|---|---|---|---|---|---|---|---|---|---|---|---|---|---|
| 1 | project | number | reference | text | document | old_codes | start_positior | end_position | data_group | text_length | word_count | doc_id | codes |
| 21116 | discern | 264:13:00 | 43 | CG020So he | sfggp_20230 | ['_cg020', '_z | 43 | 43 | ['caregiver', 'c | 1422 | 274 | discern_243( | ['caregiver', 'migration'] |
| 21117 | discern | 264:14:00 | 197 | [PERSON] IN | sfggp_20230 | ['_intv_mm', ' | 197 | 197 | ['caregiver', 'c | 64 | 10 | discern_243( | ['caregiver', 'interviewer'] |
| 21118 | discern | 264:15:00 | 443 | [PERSON] IN | sfggp_20230 | ['_intv_mm', ' | 443 | 443 | ['caregiver', 'c | 25 | 4 | discern_243( | ['caregiver', 'interviewer'] |
| 21119 | discern | 264:16:00 | 173 | CG020 INTV_ | sfggp_20230 | ['_intv_mm', ' | 173 | 173 | ['caregiver', 'c | 195 | 35 | discern_2431 | ['caregiver', 'interviewer'] |

Here is a simple example of some rows without text, opened in excel from a .csv file, from a current book project combining several large datasets.

## DATA VISUALIZATION AND ANALYSIS

Our visualizations are always used alongside in-depth quotations, to show various levels of analysis, validate typologies, support claims, or provide evidence that expected outcomes did or did not manifest (Abramson et al. 2018; Abramson et al. 2024). The tools can be used for other approaches and have been used in both scientific and humanistic settings, but our use is iterative social science, and always iterative rather than purely inductive or deductive.

Below is a sample execution block and output from an integrated development environment with Python; the current version of the code in the visualization_toolkit is linked at the top of the appendix. These visuals use n-word windows, paragraphs, full documents, or whole cases, depending on the analysis. Analysts need only understand process, parameters, and how to run the execution block to produce a wide range of visuals.

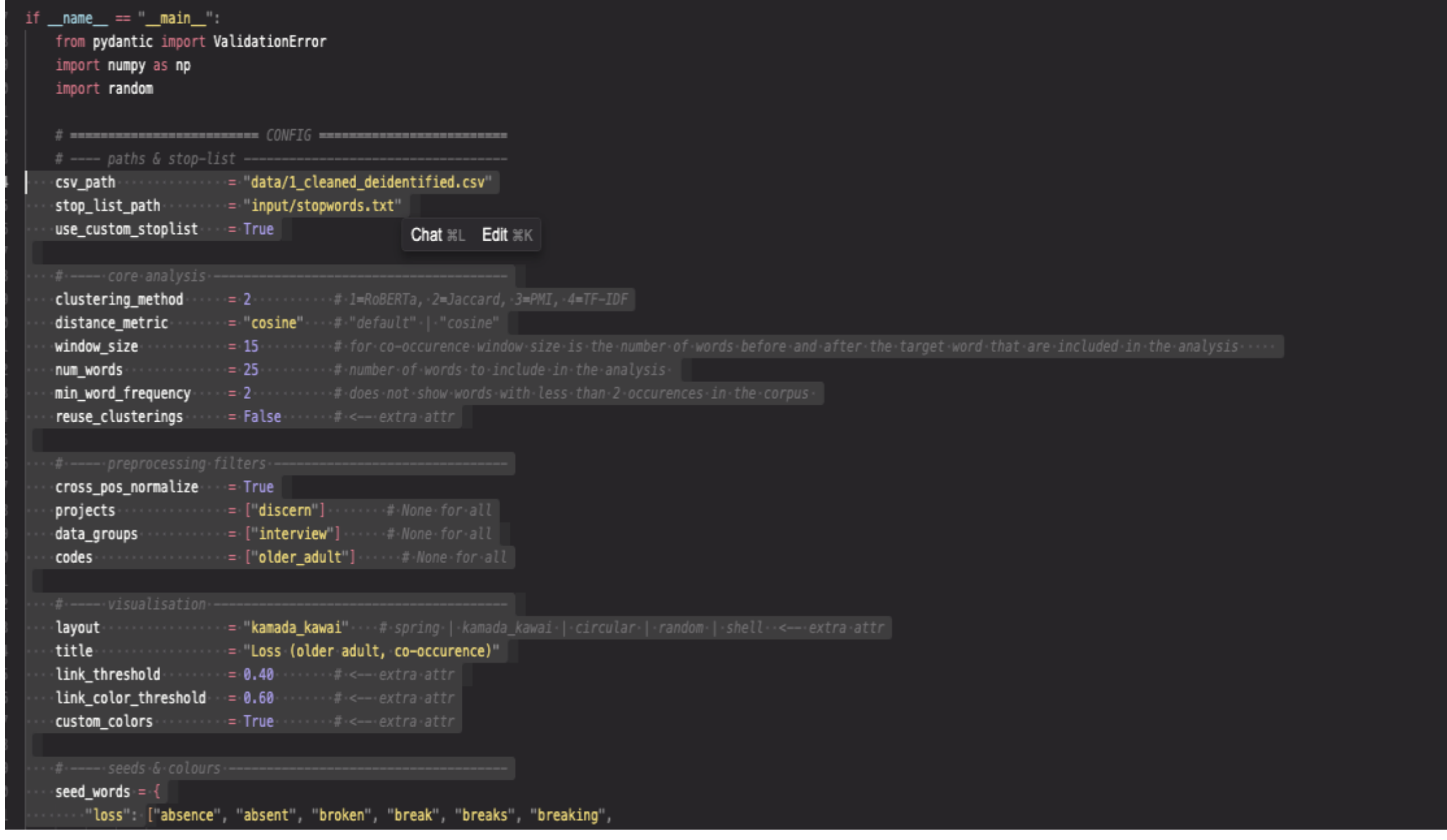

```
if __name__ == "__main__":
    from pydantic import ValidationError
    import numpy as np
    import random

    # ======================== CONFIG ========================
    # ---- paths & stop-list ------------------------------------
    csv_path              = "data/1_cleaned_deidentified.csv"
    stop_list_path        = "input/stopwords.txt"
    use_custom_stoplist   = True

    # ---- core analysis ----------------------------------------
    clustering_method     = 2          # 1=RoBERTa, 2=Jaccard, 3=PMI, 4=TF-IDF
    distance_metric       = "cosine"   # "default" | "cosine"
    window_size           = 15         # for co-occurence window size is the number of words before and after the target word that are included in the analysis
    num_words             = 25         # number of words to include in the analysis
    min_word_frequency    = 2          # does not show words with less than 2 occurences in the corpus
    reuse_clusterings     = False      # <-- extra attr

    # ---- preprocessing filters --------------------------------
    cross_pos_normalize   = True
    projects              = ["discern"]       # None for all
    data_groups           = ["interview"]     # None for all
    codes                 = ["older_adult"]   # None for all

    # ---- visualisation ----------------------------------------
    layout                = "kamada_kawai"   # spring | kamada_kawai | circular | random | shell  <-- extra attr
    title                 = "Loss (older adult, co-occurence)"
    link_threshold        = 0.40       # <-- extra attr
    link_color_threshold  = 0.60       # <-- extra attr
    custom_colors         = True       # <-- extra attr

    # ---- seeds & colours --------------------------------------
    seed_words = {
        "loss": ["absence", "absent", "broken", "break", "breaks", "breaking",
```

**Configuration block.** Standard outputs include tag- and word-based heatmaps, semantic networks (embedding or co-occurrence), dimensional reduction, simple validations, and linguistic statistics. Metadata can be integrated as filters.

The semantic network visualization shown below is from Abramson, Corey M., Turner, K., Arteaga, I., Hernández de Jesús, A., Ginn, B., Nian, Y., Dohan, D. “Pragmatic Sensemaking:

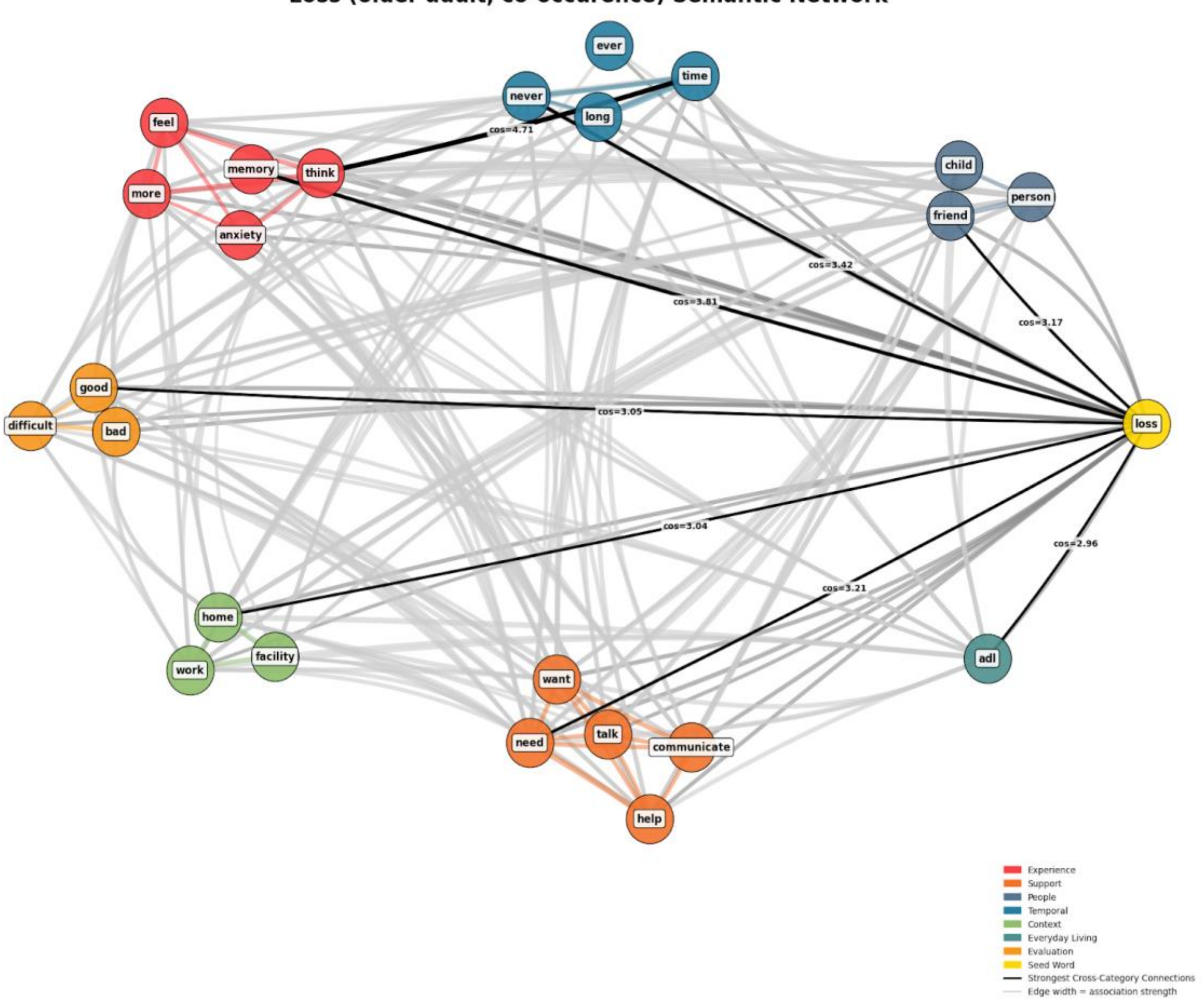

**Semantic Network**. Shows meanings of loss that people draw upon in describing life with dementia, using a cosine similarity clustered co-occurrence analysis. In the paper, this is paired with in-depth accounts to link broad language patterns and nuanced narratives.

The figure below shows an early example of an ethnoarray, subset from the larger discern study for interpretive anthropological analysis and from Arteaga 2025, in R, pre-print version [R code]. You can see an example of another ‘ethnoarray’ with historical data with Mische et al (2025) here.

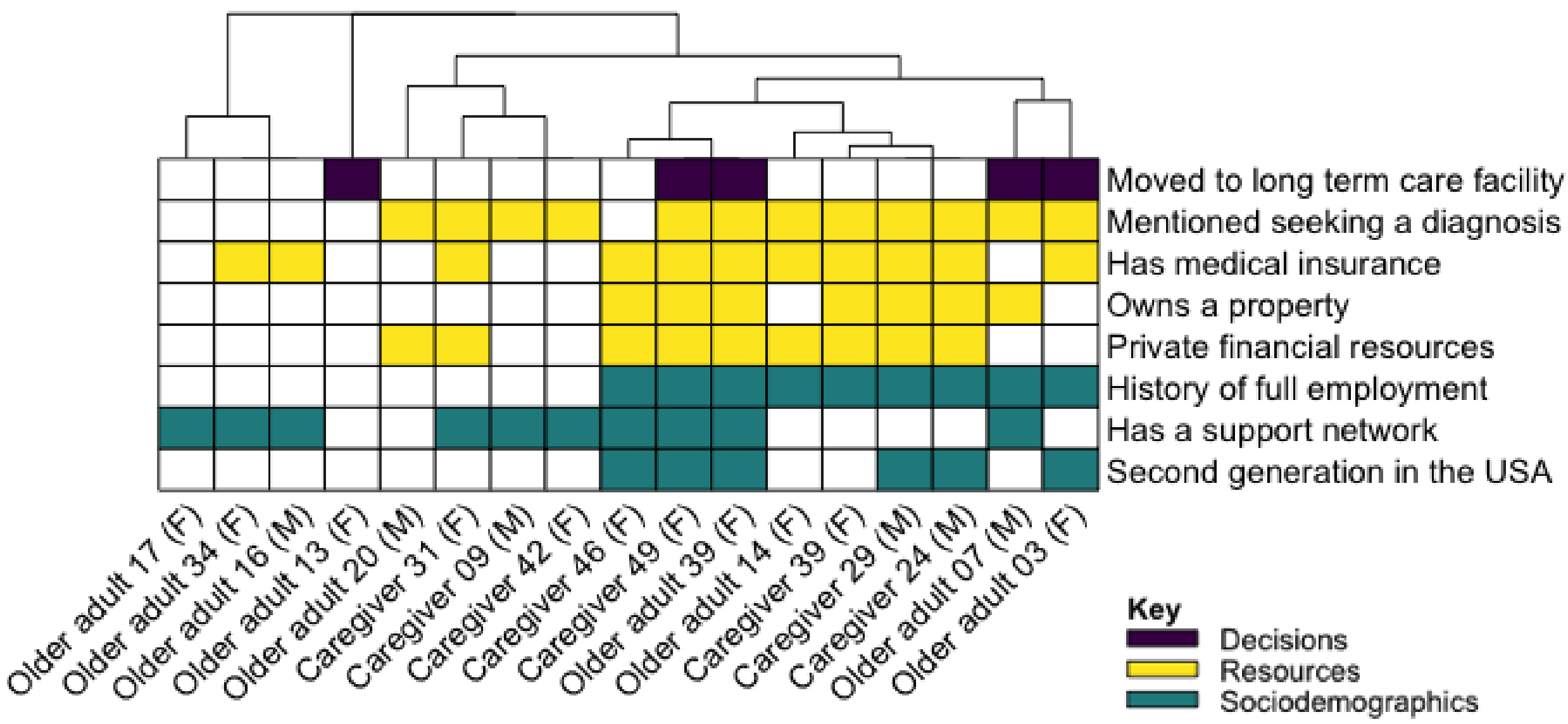

**Heatmap Case Visualization.** Showing families, resources, demographics and decisions (i.e. an ethnoarray). This was used to show variation within a group, alongside in depth ethnographic accounts. The visual is used to  help situate cases and verify clusters about the role of resources in considerations of movement to institutional settings.

## C. OTHER RESOURCES

Interactive visualization notebook, integrating heatmaps, networks, ,wordcolouds: https://colab.research.google.com/drive/1n90EDMSiXhIaOULUMPJ4W4hqdZCh1NQw?usp=sharing

For a walkthrough of a hybrid approach to coding and indexing, see "An Introduction to Machine Learning for Qualitative Research in Jupyter Notebooks (Python)." A blog description is here. American Sociological Association Methodology Workshop. (Abramson, Li, Dohan 2022). See also Li and Abramson 2025.

## D. TRANSPARENCY, VERSIONING, REPRODUCIBILITY

This article draws on tools developed for the DISCERN and prior studies, following principles of open science.  Example code and workflows are available at: https://github.com/Computational-Ethnography-Lab/teaching (version: 2025-11-01). Contents include: